# Investigation of ideal shear strength of dilute binary and ternary Ni-based alloys using first-principles calculations, CALPHAD modeling and correlation analysis

Shuang Lin[1,*], Shun-Li Shang[1,], John D. Shimanek[1], Yi Wang[1],

Allison M. Beese[1,2], and Zi-Kui Liu[1]

[1] Department of Materials Science and Engineering, The Pennsylvania State University, University Park, PA, 16802, USA

[2] Department of Mechanical Engineering, The Pennsylvania State University, University Park, PA 16802, USA

* Corresponding author: spl5745@psu.edu

**Abstract**

In the present work, the ideal shear strength ($\tau_{is}$) of dilute $Ni_{34}XZ$ ternary alloys (X or Z = Al, Co, Cr, Fe, Mn, Mo, Nb, Si, Ti) are predicted by first-principles calculations based on density functional theory (DFT) in terms of pure alias shear deformations. The $\tau_{is}$ results show that within the concentration up to 8.3% of alloying elements, $\tau_{is}$ increases with composition in binary systems with Mn, Fe, and Co in ascending order, and decreases with composition with Nb, Si, Mo, Ti, Al, and Cr in descending order. The composition dependence of $\tau_{is}$ in binary and ternary systems is modeled using the CALculation of PHAse Diagrams (CALPHAD) approach considering lattice instability, indicating that atomic bonding strength significantly influences $\tau_{is}$. Correlational analyses further show that lattice constant and elastic constant $C_{11}$ affect $\tau_{is}$, the most out of the elemental features.

**Keywords**

Ideal shear strength; First-principles calculations; CALPHAD; Correlational analysis

# 1 Introduction

The ideal shear strength of materials measures their resistance to deformation when all atomic bonds along a specific crystallographic plane are simultaneously sheared. It represents the maximum stress required for shear deformation in a defect-free crystal lattice [1]. This property can be predicted using first-principles calculations based on density functional theory (DFT). One unique advantage of DFT-based approach is its ability to be directly compared with experimental results from small-scale deformation. In these experiments, specimen geometries are carefully designed to suppress normal dislocation-mediated deformation mechanisms, enabling the applied stresses to approach the theoretical ideal strength of the material [2–4]. Leveraging DFT's intrinsic modeling capabilities, a cross-scale linkage has been recently demonstrated. For pure Ni, DFT-calculated ideal shear strengths were successfully integrated into crystal plasticity models, enabling accurate predictions of hardening behavior [5]. This also achieved by providing ideal shear strength as input data for the crystal plasticity finite element method (CPFEM), with flow resistance evaluated using the Peierls-Nabarro model [6]. This highlights ideal shear strength as a critical parameter in bridging computational and experimental studies across different scales.

In the present work, $\tau_{is}$ values of ternary $Ni_{34}XZ$ alloys are predicted through DFT-based calculations with X and Z being Al, Co, Cr, Fe, Mn, Mo, Nb, Si, and Ti, which are the major alloying elements in Ni-based Inconel alloys, for example, Inconel 625 and Inconel 718 [7,8]. These alloying elements in Ni-based superalloys play complementary roles in enhancing oxidation behavior, mechanical properties, and phase stability. Al ,Cr, Si, and Mn significantly improve oxidation resistance, with Al and Cr forming protective oxide layers, such as alumina and chromia [9,10]. While the addition of 2 at.% Si to a Ni-Al alloy is insufficient to form a protective $SiO_2$ layer, it accelerates the formation of alumina. This promotes the development of a continuous alumina scale at a relatively low temperature, further enhancing oxidation resistance above 800 °C [11]. Mn also reduces the oxidation rate without changing the scale morphology [12]. Strengthening mechanisms are primarily supported by Nb, Ti, and Mo. Nb plays a crucial role in precipitation strengthening by forming gamma double-prime ($\gamma''$) phase, which is essential for alloys like Inconel 718 [13]. Ti promotes the formation of gamma-prime ($\gamma'$) phase, further enhancing high-temperature mechanical properties [14]. Mo, as a refractory element with a high melting point, contributes significantly to solid solution strengthening [15,16]. For phase stability and mechanical performance, Co and Fe stabilize the matrix gamma ($\gamma$) phase of the superalloy [17]. Co further improves the wear resistance of the alloy [18], and Fe contributes to cost-effectiveness. Together, these elements create a synergistic balance, making Ni-based superalloys highly suitable for demanding applications.

The present study of Ni-based ternary $Ni_{34}XZ$ alloys is essential for a more comprehensive understanding of mechanical properties of multicomponent alloy systems. We aim to fill the gap in the literature by providing detailed insights into the ideal shear strength of ternary alloys which can be extrapolated to multicomponent alloys in the CALculation of PHAse Diagrams (CALPHAD) modeling framework [19], incorporating contributions from unary and binary systems, with findings that can be generalized to other multicomponent alloys. The CALPHAD method, originally developed for modeling the thermodynamic properties of multicomponent systems, has demonstrated its success in modeling phase-based properties. Its application has been extended to other properties of individual phases, for example, the diffusion coefficient (i.e., the mobility) [20], molar volume [21], elastic properties [22,23], and stacking fault energy [24]. Consequently, the general CALPHAD approach is now used as a methodology for modeling various properties as a function of the composition of individual phases. In this work, we employ the general CALPHAD approach to model the DFT-based ideal shear strengths of FCC phase as a function of composition. This approach enables the prediction of properties for unexplored compositions, providing ideal shear strength data for multicomponent Ni-based alloy databases. It supports alloy design and optimization without the need for extensive DFT calculations for every possible composition. Furthermore, the CALPHAD approach can be coupled with other modeling techniques, such as phase field simulations, to provide comprehensive insights into microstructure evolution and its impact on mechanical behavior [25].

Furthermore, correlation analyses were performed between $\tau_{is}$ and fundamental features of the alloying elements in their fcc arrangement. In our previous study, a combination of filter and wrapper methods was used, and the atomic volume and elastic properties that affected $\tau_{is}$ the most were identified [26]. In a related study, machine learning (ML) algorithms were employed to perform correlation analyses of stacking fault energy in dilute Al-, Ni-, and Pt-based alloys using elemental attributes [27]. Those analyses revealed that the variation in stacking fault energy due to alloying elements can be quantified through 14 elemental attributes, with Gaussian process regression emerging as the best ML model, exhibiting an average mean absolute error (MAE) of less than 8 mJ $m^{-2}$. The present study takes similar approaches to correlate $\tau_{is}$ with the fundamental properties of pure elements through feature selection and correlation analysis.

# 2 Methods

## 2.1 Pure alias shear deformation

As in our previous works [5,26], the pure alias shear deformation, as shown in Figure 1 (a) with one sliding plane, was adopted herein to predict the $\tau_{is}$ values using DFT-based first-principles calculations. Here, the ‘pure’ indicates full relaxations of atomic positions, cell shape, and cell volume except for the fixed shear angle. The ‘alias’ shear deformation is close to actual deformation because the displacement of one atomic layer (e.g., the top layer) influences only the next atomic layer, and the relaxations propagate from the top layer through the entire cell with fixed shear displacement [28]. Previous first-principles investigations show that deformation along the partial $\{111\}\langle 11\bar{2}\rangle$ slip system exhibits the lowest ideal shear strength for fcc metals [29].

A 36-atom orthorhombic (1×3×3) supercell was built based on a rotated six-atom fcc lattice [30] with 3 {111} layers and 12 atoms on each layer to predict $\tau_{\text{is}}$ for the $\{111\}\langle 11\bar{2}\rangle$ shear deformation. Figure 1 (b) shows this 36-atom orthorhombic cell with its lattice vectors **a, b,** and **c** along the $[11\bar{2}]$, $[\bar{1}10]$, and [111] directions, respectively. The initial lattice parameters along these three directions are set to be 8.625, 7.470, and 6.099 Å, respectively, based on the relaxed lattice parameter of pure Ni. The deformed lattice vector $\bar{\mathbf{R}}$ can be obtained by [31],

$$\bar{\mathbf{R}} = \mathbf{RD} \qquad \text{Eq. 1}$$

where **R** is the original lattice vector before alias shear, and **D** is the deformation matrix along the $[11\bar{2}]$ direction [28]:

$$\mathbf{D}_{[11\bar{2}]} = \begin{bmatrix} 1 & 0 & 0 \\ 0 & 1 & 0 \\ \varepsilon & 0 & 1 \end{bmatrix} \qquad \text{Eq. 2}$$

where ε denotes the magnitude of engineering strain, defined as the ratio between displacement applied to the supercell and the height of the supercell. The relaxations of pure alias shear deformation were performed by the external optimizer GADGET [32] to fix the shear angle during the pure alias shear deformation of $\{111\}\langle 11\bar{2}\rangle$.

## 2.2 First-principles calculations

All DFT-based first-principles calculations in the present work were performed by the Vienna Ab initio Simulation Package (VASP) [33] using the projector augmented wave (PAW) method [34]. The exchange-correlation functional was described by the generalized gradient approximation (GGA) as parameterized by Perdew et al. [35]. A $3 \times 4 \times 5$ *k*-point mesh together with a 350 eV plane wave cutoff energy was employed based on our convergence tests. The energy convergence criterion of electronic self-consistency was chosen as $2 \times 10^{-5}$ eV per supercell in all the calculations. The reciprocal space energy integration was performed by the Methfessel-Paxton technique [36] for structural relaxations with a 0.2 eV smearing width.

Due to the magnetic nature of Ni (0.66 μB/atom) [37], Co (1.7 μB/atom) [38], Fe (2.2 μB/atom) [38], Cr (4.3~4.5 μB/atom) [39], and Mn (5.0 μB/atom) [40], we examined the $\tau_{is}$ of two alloys with high magnetic moment, $Ni_{34}Cr_2$ and $Ni_{34}Mn_2$, using three kinds of spin alignments for Cr or Mn. In Table 1, the "↑↑" represents a configuration where both the spins of two Cr or Mn atoms are oriented upwards in the same direction as the spins of Ni. The "↑↓" denotes a configuration where one spin is oriented upwards and the other is oriented downwards for two Cr or Mn atoms. Lastly, the "↓↓" indicates a configuration where both spins are oriented downwards in the opposite direction of the Ni spins. Table 1 shows that the $\tau_{is}$ values differ slightly among those three spin configurations, namely by 0.017 GPa and 0.012 GPa for $Ni_{34}Cr_2$ and $Ni_{34}Mn_2$, respectively. The underlying reason is that, in all spin-down included configurations ("↑↓" and "↓↓"), spin flipping occurs during relaxations, leading to a transition to the "↑↑" configuration. As a result, these configurations inherently stabilize in the ferromagnetic (FM) configuration. This behavior is observed for the present dilute Ni-based alloys, where the magnetic interactions favor the FM state. The "↑↑" configuration was hence used in the present work.

The symmetry analysis of all configurations in terms of the positions of X and Z atoms in the supercell by the ATAT code [41] reveals 5 independent configurations (S1 to S5) when X and Z atoms are on the same (111) plane and 8 independent configurations (D1 to D8) when X and Z atoms are on different (111) planes, as shown in Figure 2. In this figure, the first alloying element X is represented by the purple sphere selected as all sites are equivalent when X is placed, while the second alloying element Z is represented by the green sphere based on the symmetry consideration. For a better representation of the various configurations, Figure 3 provides a 2D projection along the [111] direction and shows the placement of the first alloying element X, and the layer occupied by the second alloying element Z. Figure 3 (a) depicts the configuration with both X and Z on the same layer, and Figure 3 (b) with X and Z on different layers. The accompanying numbers represent the equivalent sites for the placement of the second alloying element Z.

Taking $Ni_{34}FeZ$ as an example, Figure 4 shows the relative energies for all 13 independent configurations, with the configuration S# for element Z placed at site # in Figure 3 (a) and D# for element Z placed at site # in Figure 3 (b). Figure 4 depicts that the energies of configurations S1, S2, D1, and D7 are from 0.06 to 0.34 eV/supercell, higher than those of the other 9 configurations, which range from 0 to 0.06 eV/supercell. The configuration S3 has the lowest energy for most $Ni_{34}XZ$ in the present work, except $Ni_{34}FeAl$ and $Ni_{34}FeMn$ where the D2 configuration has the lowest energy. It is noted that the configurations S3 and D2 both maximize the separation distance between Fe and Z atoms, following the maximum entropy distribution [42]. The energy differences between S3 and D2 are 0.0046 eV/supercell for $Ni_{34}FeAl$ and 0.0086 eV/supercell for $Ni_{34}FeMn$, respectively, and the differences of their ideal shear strengths are not significant, being 0.005 GPa and 0.003 GPa, respectively. Therefore, in the present work, the configuration S3 was selected to perform calculations of ideal shear strength.

## 2.3 CALPHAD modeling of ideal shear strength

In the spirit of the CALPHAD approach, $\tau_{is}$ of a multicomponent fcc solution phase can be modeled as follows [24]:

$$\tau_{is}^{fcc} = \sum_i x_i \, {}^0\tau_{i,is}^{fcc} + \sum_i x_i \sum_{j>i} x_j \, \phi_{ij}^{fcc} + \sum_i \sum_{j>i} \sum_{k>j} x_i x_j \, x_k \phi_{ijk}^{fcc} \qquad \textit{Eq. 3}$$

where ${}^0\tau_{i,is}^{fcc}$ is the ideal shear strength of pure element i in the fcc structure, $x_i$, $x_j$ and $x_k$ are the mole fractions of elements i, j, and $k$, and $\phi_{ij}^{fcc}$ and $\phi_{ijk}^{fcc}$ are the binary and ternary model parameters, respectively. For the fcc Ni-based solid solution, a fundamental challenge in *Eq.3* is how to obtain ${}^0\tau_{i,is}^{fcc}$ for all elements in the fcc structure because some elements are unstable in the fcc structure, particularly Cr, Mo, and Nb, which have the body-center-cubic (bcc) structure as their ground states [43]. This foundational topic in the CALPHAD method, i.e., the lattice stability, is an active on-going research area for unstable structures [44–46].

A recent publication [47]shows that the ideal shear strength of a pure element (${}^0\tau_{i,is}^{fcc}$) in the fcc structure can be estimated by the shear modulus in the Voigt approach, which correlates with present shear deformations. This approach shows a high correlation coefficient ($R^2$) of 0.98, as shown in Eq. 4 below

$$ {}^{0}\tau_{is}^{fcc} = 0.072 * G_{fcc}^{\{111\}<112>} \qquad R^2 = 0.98 \qquad \textit{Eq. 4} $$

The shear modulus for the fcc lattice on the $\{111\}\langle 11\bar{2}\rangle$ ($G_{fcc}^{\{111\}<112>}$) shear is derived from the elastic constants of the fcc crystal structure at their relaxed equilibrium volumes [48], as shown in Eq. 5 below

$$ G_{fcc}^{\{111\}<112>} = (C_{11} - C_{12} + C_{44})/3 \qquad \textit{Eq. 5} $$

Consequently, in present study, we apply these two equations to estimate ${}^{0}\tau_{i,is}^{fcc}$ for elements of which the ground states are not fcc. Table 2 summarizes the elastic constants of elements in fcc from a previous study [49], presenting the shear modulus and the ${}^{0}\tau_{i,is}^{fcc}$ values based on Eq. 4 and Eq. 5. In terms of elastic constants, fcc structures exhibit higher $C_{12}$ values but lower $C_{11}$ and $C_{44}$ values compared to their ground states. For example, the $C_{11}$ and $C_{44}$ values for fcc-Cr are 110.7 GPa and -36.5 GPa, respectively, significantly lower than the bcc-Cr values of 247.6 GPa and 48.3 GPa. The negative elastic constant values shown in fcc-Cr ($C_{44}$) and fcc-Nb ($C_{11}$ and $C_{44}$) represent their instability, which also leads to negative values in both the shear modulus and ideal shear strength, particularly in elements like Cr, Mo, Si, and Nb.

Additionally, Table 2 presents the ideal shear strength of each pure element in their ground state $\tau_{is}^{gs}$. It is shown that elements with a hexagonal close-packed (hcp) ground state, such as Co and Ti, exhibit an ideal shear strength more closely aligned with their $\tau_{is}^{fcc}$ values, compared to elements whose ground state is bcc (Fe, Mn, Cr, Mo, Nb) and that with a diamond structure (Si).

This finding is understandable, as the close-packed planes in fcc and hcp are identical. This study successfully represents ${}^{0}\tau_{i,is}^{fcc}$ for elements whose ground states are not fcc by using the fcc elastic constant and shear modulus as shown by Eq. 4 and Eq. 5 enabling the determination of $\phi_{ij}^{fcc}$ and $\phi_{ijk}^{fcc}$ within the full composition range using the general CALPHAD modeling approach.

## 2.4 Correlation analysis and feature selection

To gain insights to the relations between physical features of alloying elements and the $\tau_{is}$ values, the variation of $\tau_{is}$ was examined by correlation analysis. The present work investigates the correlation between the $\tau_{is}$ values of $Ni_{34}XZ$ and the linearly combined features, i.e., the average feature values of those of X and Z by considering the equal mole fractions of X and Z in $Ni_{34}XZ$. The major physical features were chosen based on our previous work on the effects of alloying elements on $\tau_{is}$ of 26 $Ni_{11}X$ binary alloys [26]. They are grouped in Table S3 in several categories of elemental attributes including atomic, periodic, elastic, thermodynamic, lattice, and electronic properties [27]. Additionally, compared with previous studies the present work also investigates the effects of features from first-principles predictions, including elastic constants ($C_{ij}$) and lattice constants. The $C_{ij}$ values for feature analysis were based on calculations of metals with fcc structure [49,50].

Feature selection methods [51]can be categorized into filter type and wrapper type methods [52]. Filter type methods are based on statistical tests of correlation, including the coefficient of determination of linear fitting (i.e., $R^2$), the maximal information coefficient (MIC), F-tests, and feature scores from a regression relief algorithm. $R^2$ represents the proportion of variance in the dependent variable explained by the independent variable, ranging from 0 to 1, with higher values indicating a stronger linear correlation. The MIC evaluates the non-linear correlation between two variables, conceptualized by a partitioning grid separating the scatterplot of two variables [32]. The F-tests statistically examine the variance of two variables by extending the $R^2$ method. The relief algorithm quantifies correlations by adjusting feature weights based on their ability to explain the variance of targeted features [53]. In wrapper type methods, a subset of features will be used to train a machine learning model. A metric of model performance, for example, the mean squared error (MSE), will determine whether features are added into or removed from the feature subset. Both the forward and the backward feature selection modes can be used in the wrapper methods. In the forward scheme, features are sequentially added into a null set, and in the backward scheme, features are sequentially removed from a full-feature set.

All the filter type and wrapper type methods used in the present work are coded in MATLAB (version R2020b) [54]. A rational quadratic gaussian process regression (GPR) model was used in the wrapper method based on our tests (see refs. [26,55]) for the forward and backward feature selections, with the 5-fold cross validation for determining model performance in terms of the MSE value. A total of 1000 iterations of the wrapper method were used in the present work to ensure convergence of feature selections.

# 3 Results and discussion

## 3.1 First-principles calculations

During deformation, the shear stress component initially increases with applied strain, followed by a sudden drop upon reaching a maximum value. This point of the maximum stress just before the drop is considered as $\tau_{is}$ of the alloy. Figure 5 (a) illustrates a representative example ($Ni_{34}CoNb$) of this shear stress behavior with the total energy also plotted. The highest stress value corresponds to the point on the total energy curve with the steepest slope, as enlarged in Figure 5 (b), since stress can be defined as the negative derivative of total energy with strain. When the total energy subsequently reaches its peak value at the unstable stacking fault point, the stress becomes zero. However, not all alloys exhibit such a sharp drop in stress. Figure 6 illustrates a gradual drop-off in stress for the $Ni_{34}CoNb$ and $Ni_{34}NbNb$ cases, often in systems with lower $\tau_{is}$, such as alloys containing Nb with a low bond strength.

The calculated ideal shear strength of pure fcc Ni is ${}^{0}\tau_{Ni,is}^{fcc} = 5.09$ GPa using the present 36-atom supercell, which is in good agreement with the previous pure alias shear studies (5.0 GPa using a 12-atom supercell [56] and 5.13 GPa using a 6-atom supercell [26]). Table 3 lists ${}^{0}\tau_{Ni,is}^{fcc}$ and the corresponding engineering shear strains for various supercell sizes. The engineering strain remains constant at 0.12 for the 6-, 12-, and 36-atom supercells. The shear strength of pure fcc Ni measured by nanoindentation measurement is 8 GPa [2]. The difference in $\tau_{is}$ values between DFT-based calculations and experimental measurements can be attributed to complex stress states during the nanoindentation process in comparison with the simple ones in the DFT-based calculations and the difference of temperature [4,34].

Table 4 summarizes the predicted $\tau_{is}^{fcc}$ values of $Ni_{34}XZ$ from the DFT-based calculations in the present work, which are in the range of 4.16 to 5.24 GPa, with the highest $\tau_{is}$ for $Ni_{34}CoCo$ and the lowest for $Ni_{34}NbNb$. It is observed that Mn, Fe, and Co increase $\tau_{is}$ in the ascending order, while Nb, Si, Mo, Ti, Al and Cr decrease $\tau_{is}$ in the descending order. Stronger atomic bonding, often associated with small atomic volume [57], can lead to higher ideal shear strength, as seen with Co, Mn, and Fe in the present work and in agreement with our previous findings [26].

Furthermore, the magnetic nature of Co, Mn and Fe contributes to directional bonding and increases the bond strength [58]. The density of valence electron charge also plays a crucial role in bonding: the denser the electron charge, the stronger the bond between atoms [31]. Elements such as Co ($3d^7 4s^2$), Fe ($3d^6 4s^2$), and Mn ($3d^5 4s^2$) display higher valence electron densities compared to Nb ($5s^1 4d^4$), Si ($3s^2 3p^2$), Mo ($5s^1 4d^5$), Ti ($4s^2 3d^2$), Al ($3s^2 3p^1$), and Cr ($3d^5 4s^1$).

Furthermore, the ideal shear stress of $Ni_{34}XZ$ can be estimated by $Ni_{34}X_2$ and $Ni_{34}Z_2$ based on the rule of mixture (ROM) principle, written as follows:

$$\tau_{\mathrm{is}}^{\mathrm{ROM}}(\mathrm{Ni}_{34}\mathrm{XZ}) = 0.5 * \tau_{\mathrm{is}}(\mathrm{Ni}_{34}\mathrm{X}_2) + 0.5 * \tau_{\mathrm{is}}(\mathrm{Ni}_{34}\mathrm{Z}_2) \quad \textit{Eq. 6}$$

with $\tau_{\mathrm{is}}(\mathrm{Ni}_{34}\mathrm{X}_2)$ and $\tau_{\mathrm{is}}(\mathrm{Ni}_{34}\mathrm{Z}_2)$ being the ideal shear stresses of $Ni_{34}X_2$ and $Ni_{34}Z_2$, respectively. A difference between $\tau_{is}^{ternary}$ from the DFT-based calculations and $\tau_{\mathrm{is}}^{\mathrm{ROM}}$ would indicate that nonlinear effects exist that are not captured in Eq. 6. However, Figure 7 shows a strong correlation between $\tau_{is}^{ternary}$ and $\tau_{\mathrm{is}}^{\mathrm{ROM}}$ with a coefficient of determination ($R^2 = 0.971$) and a low mean absolute error (MAE) of 0.03 GPa. This suggests that in the small composition range of most ternary alloys in the present work, the ideal shear strength can be effectively estimated using Eq. 6. This is particularly true for $Ni_{34}AlCo$, $Ni_{34}AlFe$, $Ni_{34}AlTi$, $Ni_{34}AlMo$, and $Ni_{34}AlSi$, which show an absolute difference of within 0.01 GPa between $\tau_{is}^{ternary}$ and $\tau_{\mathrm{is}}^{\mathrm{ROM}}$ (see Table S1). This linearity can be explained as Al has the same fcc ground state structure as the host element Ni. However, the nonlinearity increases in alloys like $Ni_{34}AlNb$, $Ni_{34}AlMn$ and $Ni_{34}AlCr$, which exhibit an absolute difference up to 0.075 GPa, probably due to the change in the magnetic moment during shear deformation, as shown in Table S1.

### 3.2 CALPHAD modeling of ideal shear strength $\tau_{\mathrm{is}}$

All the binary interaction parameters ${}^{0}\phi_{Ni,j}^{fcc}$ and ternary interaction parameters ${}^{0}\phi_{Ni,j,k}^{fcc}$ were evaluated using Eq. 3, with $\tau_{is}$ values from first-principles calculations. The ideal shear strength of pure elements whose ground states are not fcc was approximated by Eq. 4, as discussed in section 2.3, with values given in Table 2.

Specifically, ${}^{0}\phi_{Ni,j}^{fcc}$ was modeled with $\tau_{is}$ of $Ni_{35}X$, $Ni_{34}X_2$ from the present work and $Ni_{11}X$ from our previous work (Table S2) [26]. It is worth mentioning that there is only one independent configuration of $Ni_{35}X$ or $Ni_{11}X$ when one alloying element X on the top layer of the supercell. For $Ni_{34}X_2$, the lowest configuration S3 was used, as determined in section 2.2. Table 5 shows the values of ${}^{0}\phi_{\mathrm{Ni,j}}^{\mathrm{fcc}}$ for each element. Figure 8 plots the fitting result of ${}^{0}\phi_{\mathrm{Ni,j}}^{\mathrm{fcc}}$ in Ni-X with alloying concentration from 0-10%, with details for each element shown in supplementary file, Figure S1. As Figure 8 shows, at low concentration ranges, the ideal shear strength varies approximately linearly with the concentration. In Figure S1, we compared the Redlich–Kister (R-K) [59] polynomial fit for all elements with a linear fit, excluding the pure end points of the alloying elements. The $R^2$ results indicate that the R-K fit is the better fit for the low concentration region. It should be mentioned that the foundational concept of the CALPHAD modeling is the lattice stability, i.e., the property of a pure element in structures other than its stable structure at room temperature [46,60]. The $R^2$ for most alloying elements is excellent, ranging from 0.99 to 1.00, except Mn with a slightly lower $R^2$ value of 0.928, in the concentration range, i.e., up to 8.3%. It is also necessary to point that ${}^{0}\phi_{\mathrm{Ni,j}}^{\mathrm{fcc}}$ carries uncertainties due to the missing data points beyond the Ni-rich region and the shortcomings of polynomial fitting of interaction parameters. Consequently, the uncertainty to predict ideal shear strength using ${}^{0}\phi_{\mathrm{Ni,j}}^{\mathrm{fcc}}$ increases once the alloying concentration exceeds 8.3%. The fits for all alloying elements have their p-values below the 5% significance threshold, indicating that the fitted models for these alloying elements are statistically significant.

The ternary interaction parameter ${}^{0}\phi^{fcc}_{Ni,j,k}$ was evaluated using $\tau_{is}$ values from $Ni_{34}XZ$, with the fixed binary interaction ${}^{0}\phi^{fcc}_{Ni,X}$ and ${}^{0}\phi^{fcc}_{Ni,Z}$ as determined previously. The binary interaction between X and Z, ${}^{0}\phi^{fcc}_{X,Z}$ in Eq. 3, was ignored due to their low concentrations. Table 6 presents the values of ${}^{0}\phi^{fcc}_{Ni,j,k}$. For most cases, ${}^{0}\phi^{fcc}_{Ni,j,k}$ are negative, which means $\tau_{is}$ decreases due to ternary interaction, particularly with the presence of Mo and Nb due to their larger atomic volumes and low bonding strength.

3.3 Correlational analyses based on elemental features

Several feature selection techniques were utilized to understand the underlying physics that governs the relations between $\tau_{is}$ and the properties of alloying elements X and Z. The 47 features of alloying elements presented in Table S3 are ranked in terms of their importance determined by filter and wrapper approaches, and the results are listed in Table S4. Although each approach assesses the importance of a feature differently, there is a consistent trend observed in the feature rankings across various selection methods as discussed below.

Figure 9 (a) reveals a linear correlation between $\tau_{is}$ and lattice constant with an $R^2$ value of 0.820, showing larger lattice constant of alloying elements corresponding to lower values of $\tau_{is}$. This can be explained by a larger lattice constant, indicating that atoms are farther apart, resulting in a weaker atomic bond strength. The reduced interatomic strength mean that less stress is required to initiate the sliding of one plane of atoms over another, leading to a lower ideal shear strength.

Figure 9 (b) plots $\tau_{is}$ versus the elastic constant $C_{11}$ with an $R^2$ value of 0.812, suggesting that a larger $\tau_{is}$ corresponds to a higher $C_{11}$ value, indicating stronger bonds within the crystal lattice, which enable the material to resist greater shear forces, thereby increasing $\tau_{is}$. Meanwhile, some attributes like electronegativity on the Pauling scale, consistently ranked low, indicating they are less influential on $\tau_{is}$ (see Table S4).

Figure 10 provides a visual representation of the five most important features, their specific scores from F-test, Relief, $R^2$, and MIC evaluations, and their occurrence frequencies as determined by forward and backward feature selection. In addition to lattice constant and elastic constant $C_{11}$, atomic volume and cohesive energy consistently rank high in the filter method, suggesting a significant correlation between the $\tau_{is}$ and features related to bonding strength. Specifically, a smaller atomic volume and higher cohesive energy are associated with stronger bonding, which in turn leads to a higher $\tau_{is}$. The feature describing lattice constant consistently ranks high in all methods, except for the backward method. In the forward and backward feature selection methods, the Debye temperature stands out as an important feature, ranking second in the forward method and first in the backward method. This prioritization of Debye temperature is understandable as it reflects the stiffness of the atomic lattice; a higher Debye temperature represents a stiffer lattice with more robust atomic bonds and higher vibrational frequencies. The literature further supports this by showing that Debye temperature can be estimated from the elastic moduli [61,62], which also supports that elastic properties of the elements have the significant effect on $\tau_{is}$. The similarity in the most important features between the present ternary systems and the previous binary systems suggests resemblance in their chemical influence on deformation, but a more quantitively detailed accounting of the elemental interactions is found in the CALPHAD framework for ideal shear strength. It should be mentioned that the wrapper method results in Table S4 may not fully account for critical information identified by the sequential algorithm. This is because if two features carry the same key information, they may be alternately selected across iterations, each appearing with only a half frequency in the final count, even though the information they represent is considered fully of the cases. Therefore, features with overlapping but critical information relevant to the shearing process might receive a lower frequency ranking compared to moderately related features [26].

## 4 Summary

In the present work, $\tau_{is}$ of dilute ternary $Ni_{34}XZ$ alloys are predicted by DFT-based first-principles calculations with the pure alias shear deformation. Additions of Mn, Fe, and Co are shown to enhance $\tau_{is}$ in ascending order, while Nb, Si, Mo, Ti, Al, and Cr decrease it in descending order. The present study demonstrates that $\tau_{is}$ is governed by the strength of atomic bonds, which are influenced by factors such as atomic volume, valence electron concentration, and the magnetic properties of the elements. There is a strong linear correlation between $\tau_{is}$ of ternary systems and the weighted sum of $\tau_{is}$ of constitutive binary systems, but nonlinearity increases when there is a significant change in magnetic moment of the alloys during shear deformation. The composition dependence of $\tau_{is}$ is modeled using the general CALPHAD method, with inputs from DFT-based $\tau_{is}$ calculations. CALPHAD approach enables the prediction of $\tau_{is}$ across the entire compositional range. Feature selection methods, both filter-type and wrapper-type, consistently show that the lattice constant and the elastic constant $C_{11}$ are critical elemental features for predicting $\tau_{is}$; a larger lattice constant correlates with a lower $\tau_{is}$, while a higher elastic constant is associated with a higher $\tau_{is}$.

## 5 cRediT authorship contribution statement

Shuang Lin: Investigation, Formal analysis, Data curation, Writing – original draft, Writing – review & editing. Shun-Li Shang: Conceptualization, Methodology, Supervision, review & editing, Funding acquisition. John D. Shimanek: Conceptualization, review & editing. Yi Wang: Conceptualization, Methodology. Allison M. Beese: Conceptualization, Project administration, Resources, Writing – review & editing, Funding acquisition. Zi-Kui Liu: Conceptualization, Project administration, supervising, Resources, Writing – review & editing, Funding acquisition.

## 6 Declaration of Competing Interest

The authors declare that they have no known competing financial interests or personal relationships that could have appeared to influence the work reported in this paper.

## 7 Acknowledgements

The present work was financially supported by the U. S. Department of Energy (DOE) via award nos. DE-FE0031553 and DE-AR0001435. First-principles calculations were performed partially on the Roar supercomputer at the Pennsylvania State University's Institute for Computational and Data Sciences (ICDS), partially on the resources of the National Energy Research Scientific Computing Center (NERSC) supported by the U.S. DOE Office of Science User Facility operated under Contract No. DE-AC02-05CH11231, and partially on the resources of the Extreme Science and Engineering Discovery Environment (XSEDE) supported by NSF with Grant No. ACI-1548562

## 8 Data availability

Data generated and analyzed in the present study are available from the corresponding author upon request.

## 9 Figures

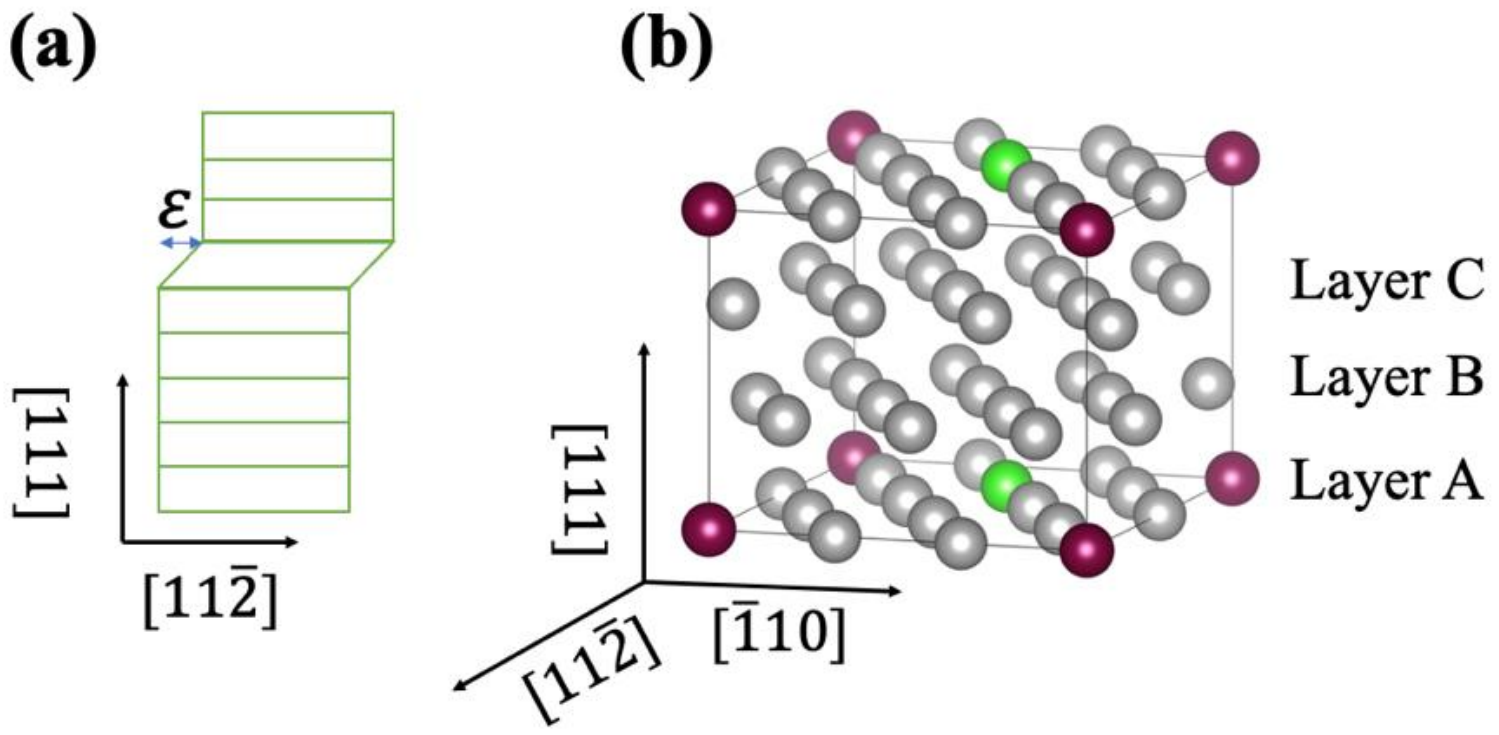

*Figure 1. Schematic representations of (a) the alias shear deformation, where ε is the magnitude of shear strain, and (b) the orthorhombic FCC lattice with 36-atom supercell, showing Ni atom (gray) and alloying atoms X (purple) and Z (green) with three {111} layers and 12 atoms on each layer are employed.*

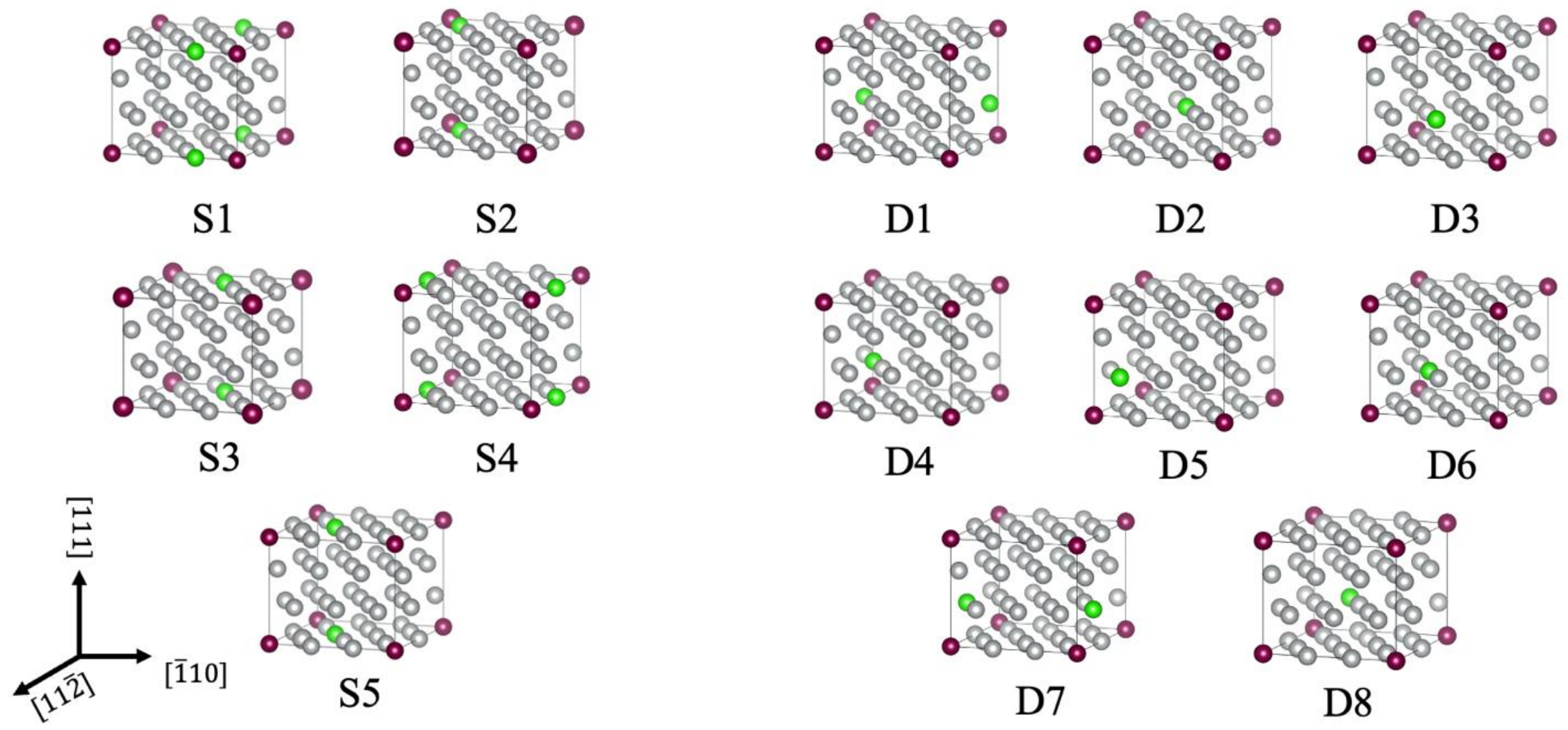

*Figure 2. Schematic representations of independent configurations with the first alloying element X denoted by purple atom on the corner sites and the second alloying element Z demoted by green atom. S1 to S5 represents when X and Z atoms are on the same (111) plane, and D1 to D8 represent when X and Z atoms are on different (111) planes.*

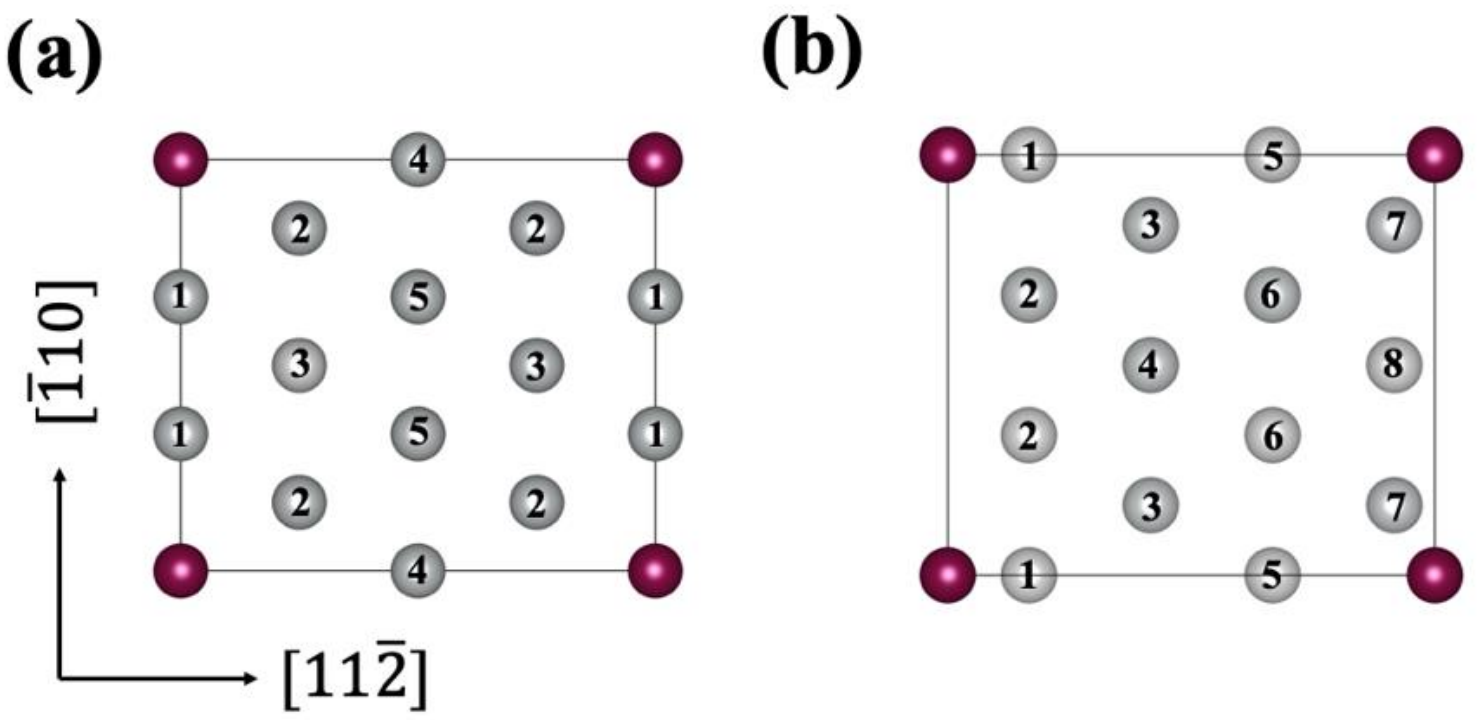

*Figure 3. Projection view of supercell along [111] direction. (a) shows scenarios where the alloying elements X and Z are on the same layer, while (b) shows those where X and Z are positioned on different layers. X is denoted by purple atom and the accompanying numbers indicate equivalent sites for the placement of the second alloying element Z.*

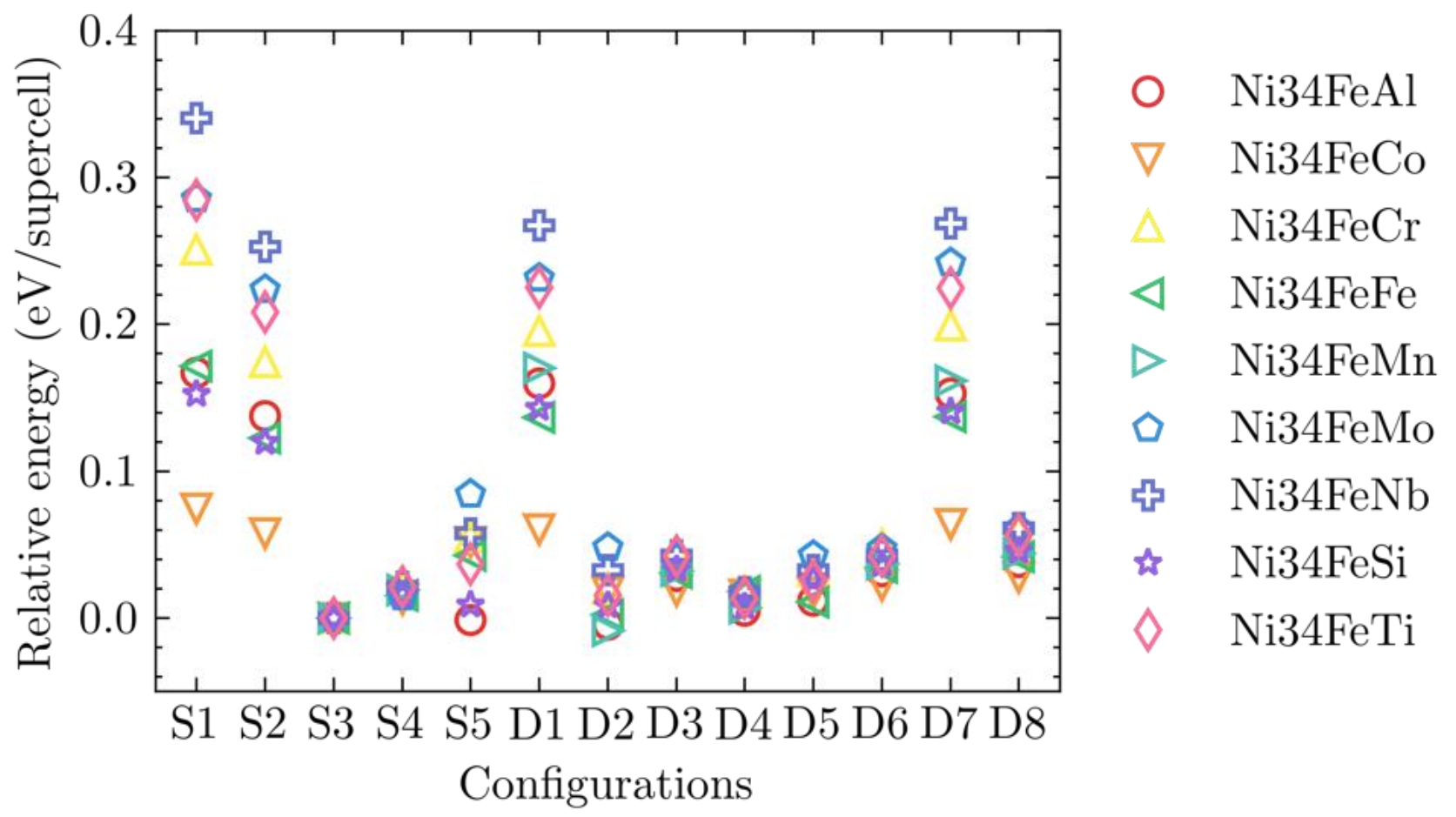

*Figure 4. Relative energies of the 13 independent structures for $Ni_{34}FeZ$ in terms of S1 to S5 and D1 to D8 configurations. The configuration S3 is one of the lowest energy configurations for most Z elements which is selected to perform first-principles calculation.*

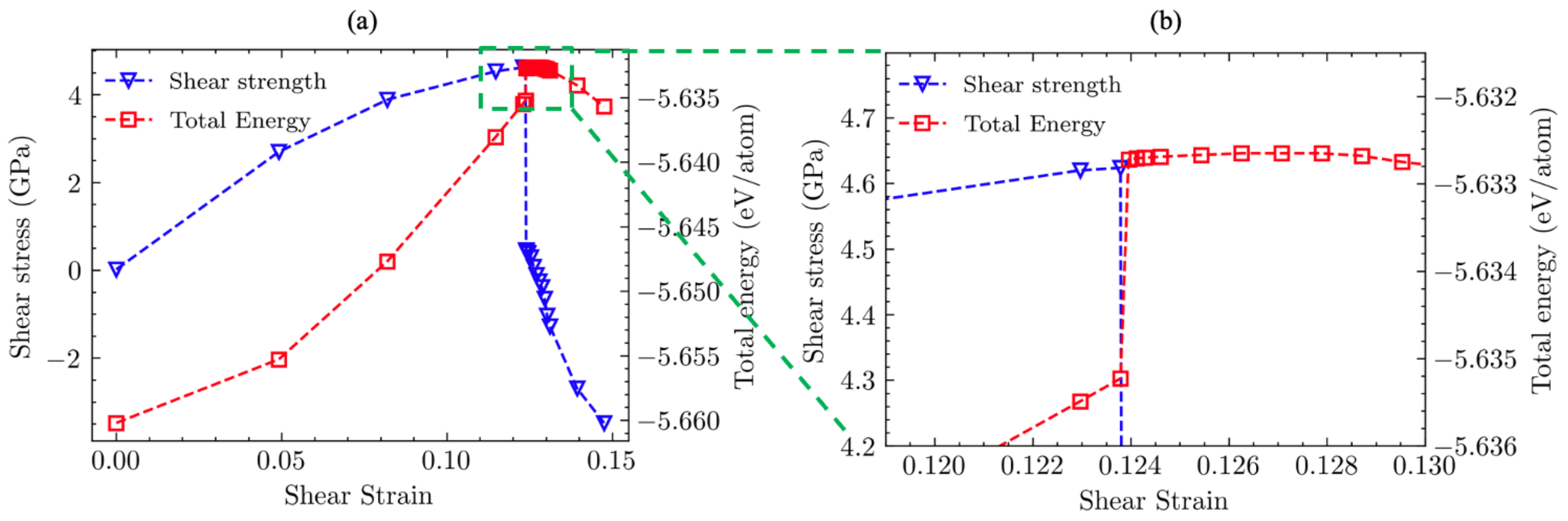

*Figure 5.(a) Changes of total energy and shear stress with respect to shear strain for Ni34CoNb by first-principles calculations. It shows the increase of shear stress with increasing shear strain before reaching the maximum point, i.e., the ideal shear strength ($\tau_{is}$), and the slope of total energy with respect to shear strain continuously increases when approaching $\tau_{is}$. (b) Zoomed in plot shows that the $\tau_{is}$ corresponds to the jump of total energy.*

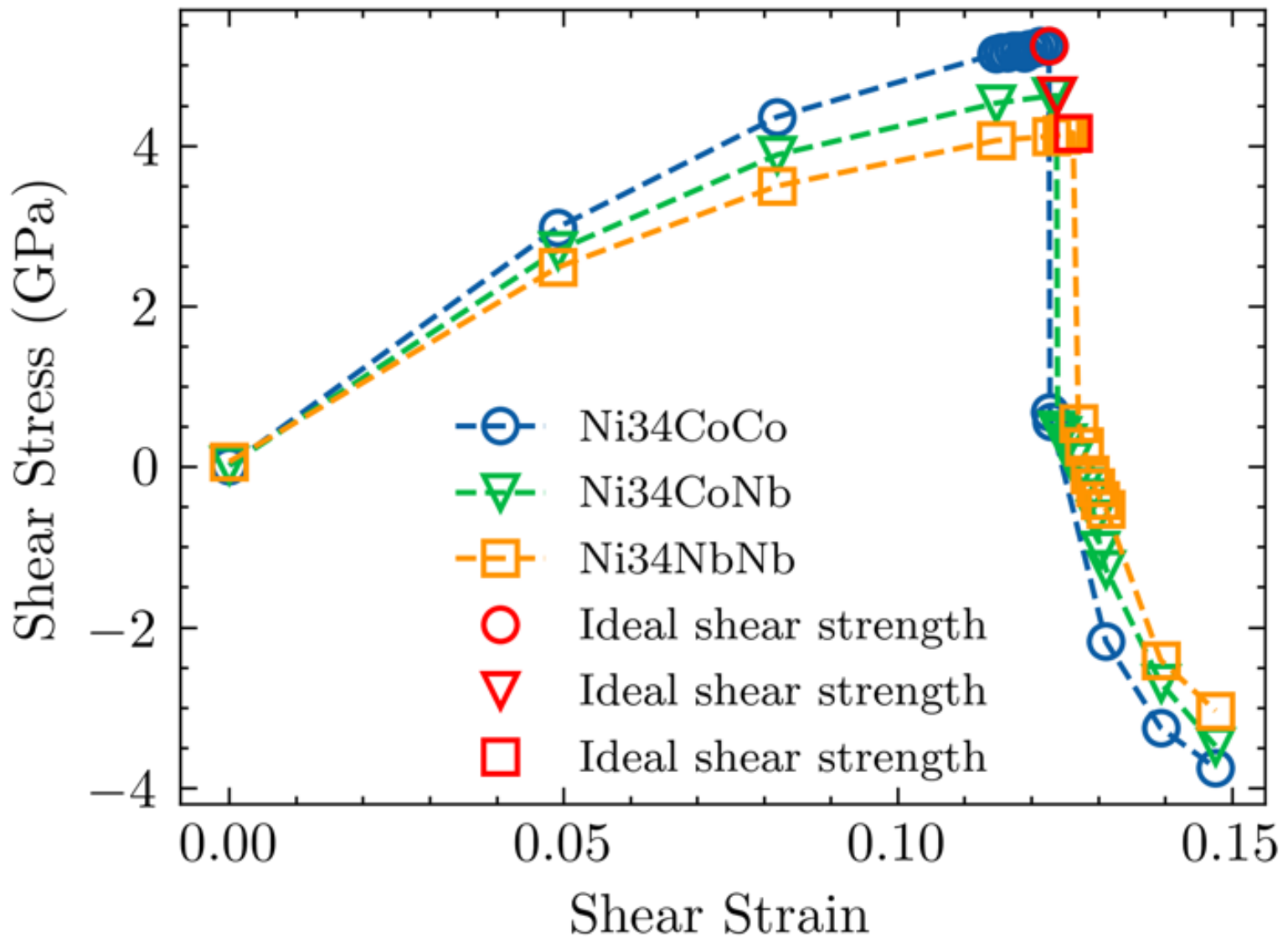

*Figure 6. Calculated shear stresses vs shear strains for $Ni_{34}CoCo$, $Ni_{34}CoNb$ and $Ni_{34}NbNb$. The red symbols represent the highest values of shear deformations, which are the $\tau_{is}$. A higher $\tau_{is}$ is corresponding to a lower shear strain*

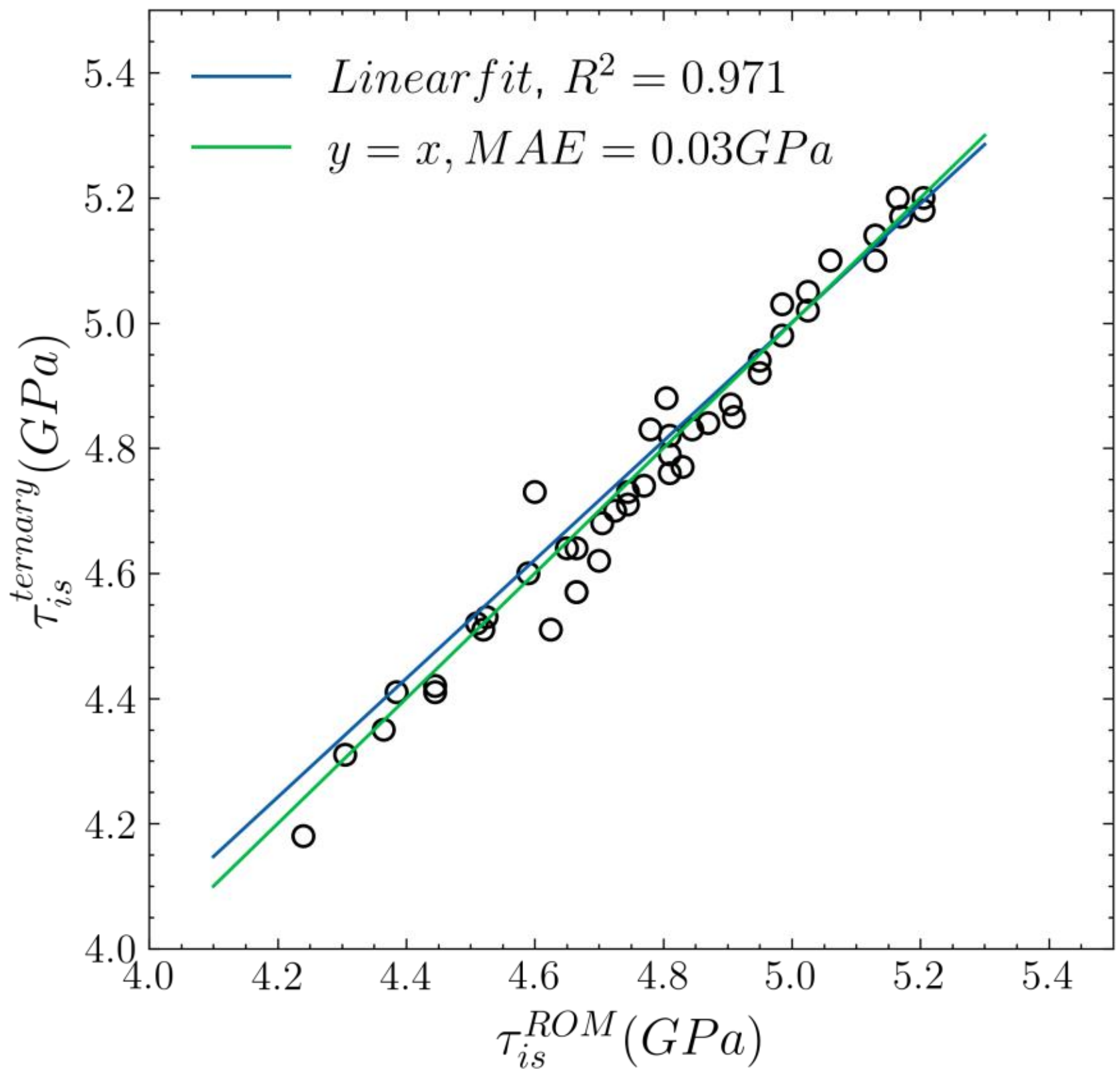

*Figure 7. Correlation between $\tau_{is}^{ternary}$ of ternaries ($Ni_{34}XZ$) and $\tau_{is}^{ROM}$ from values of binaries ($Ni_{34}X_2$) along with the linear fit line with $R^2$=0.971, and a reference line y=x exhibiting a mean absolute error (MAE) of 0.03 GPa.*

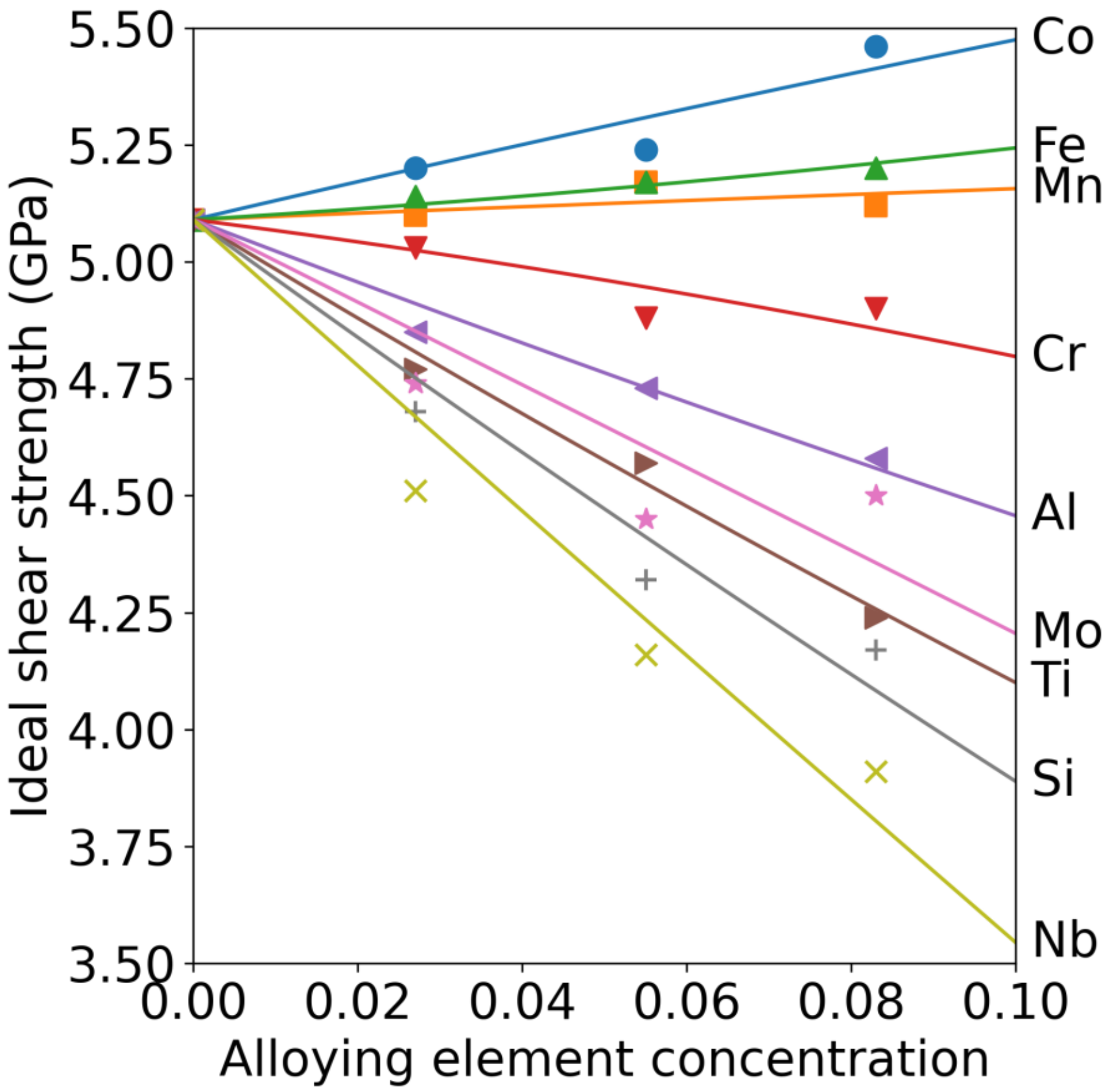

*Figure 8. Ideal shear strength as a function of alloying element concentration. The data of Ni Ni$_{35}$X, Ni$_{34}$X$_2$ from the present work as well as those of Ni$_{11}$X are also labeled in the plot.*

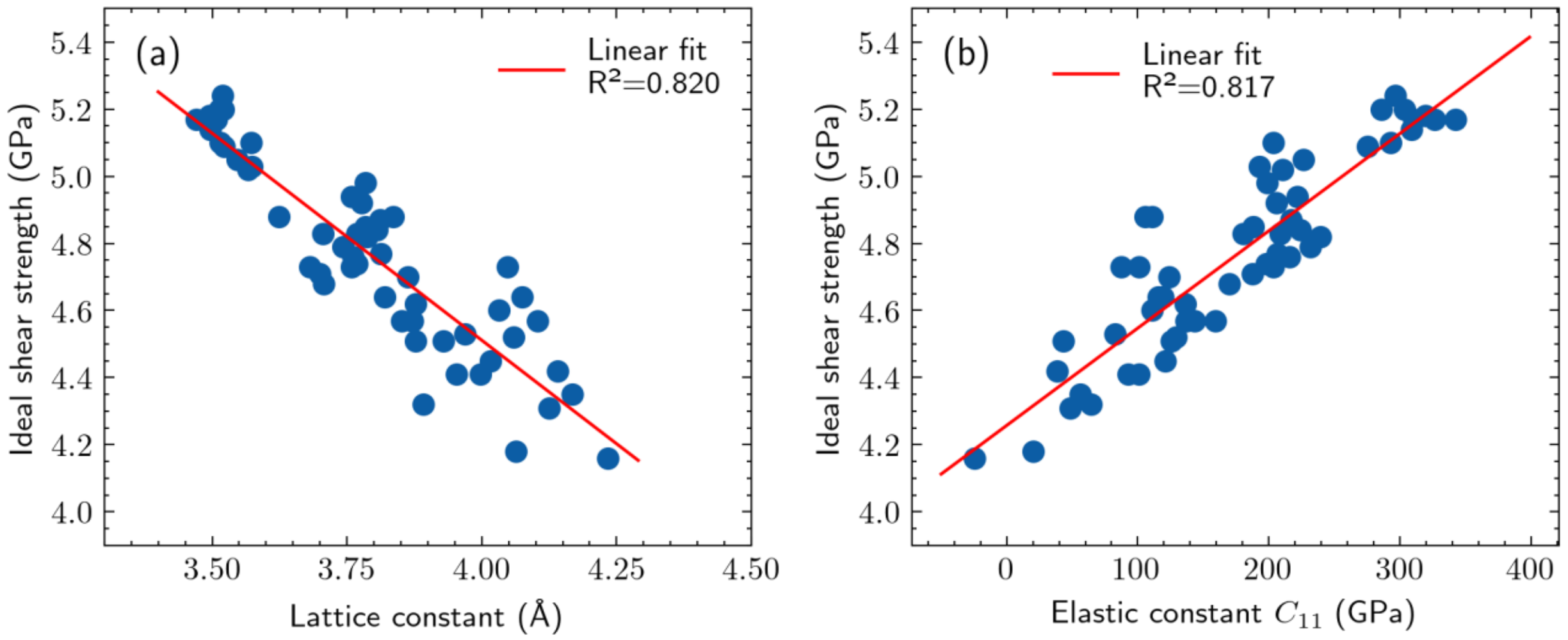

*Figure 9. (a) Ideal shear strength as a function of the linearly combined DFT-calculated lattice parameter of element X and Z with linear fitting $R^2$= 0.820. (b) Ideal shear strength as a function of the linearly combined elastic constant $C_{11}$ of alloying elements X and Z with linear fitting $R^2$= 0.817.*

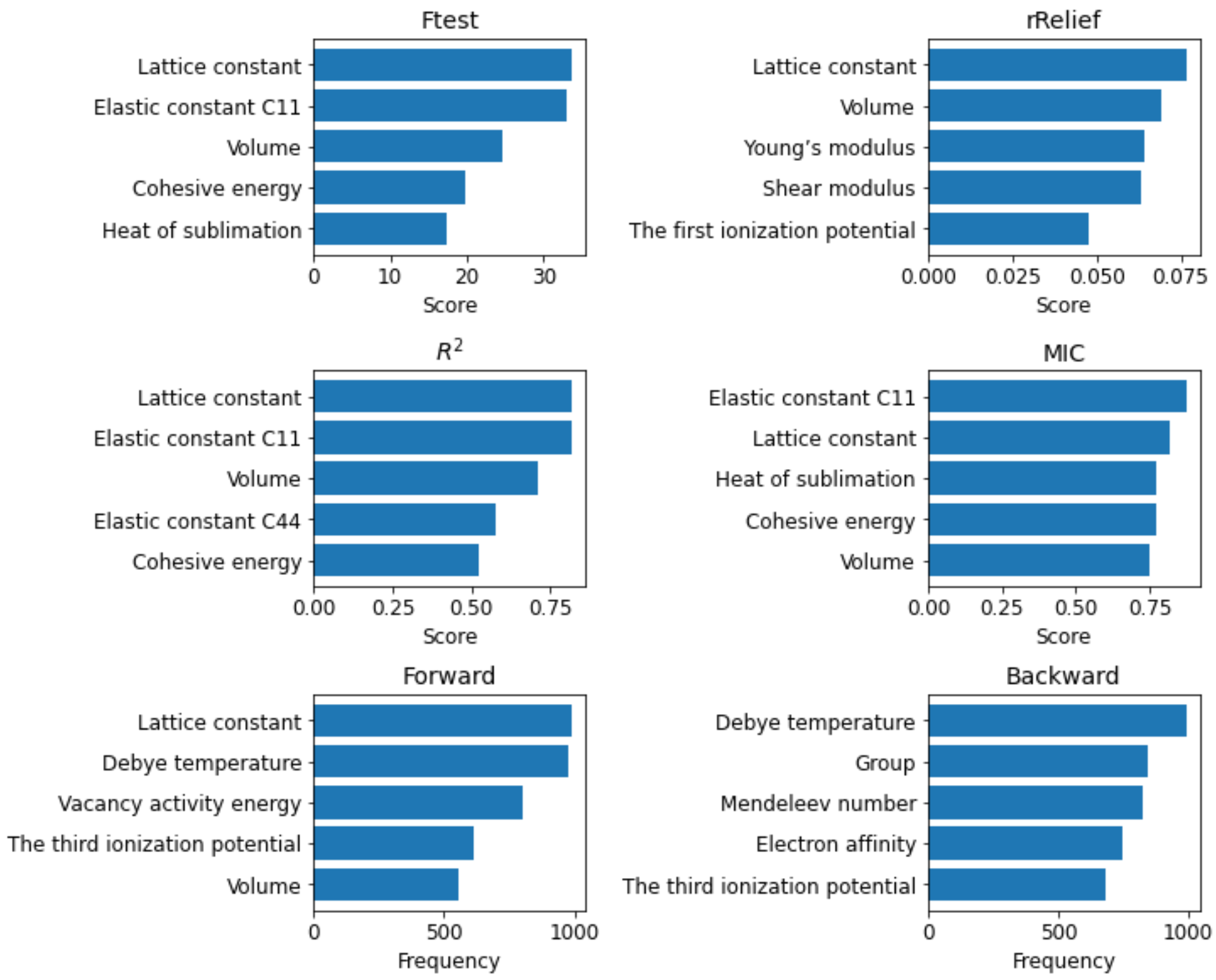

*Figure 10. Rankings of the top 5 features as determined by four filter selection methods (Ftest, rRelief, R_squared, and MIC) and two wrapper selection methods (forward and backward). The filter methods consistently agree the lattice constant and elastic constant $C_{11}$ as important features, while the forward and backward feature selection underscore the Debye temperature is important feature.*

## 10 Tables

*Table 1. Ideal shear strength $\tau_{is}$ prediction of $Ni_{34}Cr_2$ and $Ni_{34}Mn_2$ under three different electron spin-polarizations direction of Cr or Mn. In this table, ↑↑ denotes the arrangement where both spins of two Cr atoms in $Ni_{34}Cr_2$ or two Mn atoms in $Ni_{34}Mn_2$ are oriented upwards. ↑↓ denotes a configuration where one spin is oriented up while the other is oriented down. Conversely, ↓↓ denotes a state where both spins are oriented down.*

| Alloys | Electron spin-polarization directions of Cr or Mn | Ideal shear strength $\tau_{is}$ (GPa) |
|---|---|---|
| $Ni_{34}Cr_2$ | ↑↑ | 4.879 |
| | ↑↓ | 4.868 |
| | ↓↓ | 4.885 |
| $Ni_{34}Mn_2$ | ↑↑ | 5.175 |
| | ↑↓ | 5.187 |
| | ↓↓ | 5.182 |

*Table 2. Elastic Constants ($C_{11}$, $C_{12}$ and $C_{44}$) from first-principles calculations in fcc structure* [49], *with $G_{fcc}^{\{111\}<112>}$ and $\tau_{is}^{fcc}$ derived from Equations 4 and 5, and $\tau_{is}^{gs}$ representing ideal shear strength in their ground states [47].*

| Elements | $C_{11}$ (GPa) | $C_{12}$ (GPa) | $C_{44}$ (GPa) | $G_{fcc}^{\{111\}<112>}$ (GPa) | $\tau_{is}^{fcc}$ (GPa) | $\tau_{is}^{gs}$ (GPa) |
|---|---|---|---|---|---|---|
| Co | 296.59 | 171.88 | 143.98 | 89.57 | 6.45 | 6.55 |
| Fe | 342.25 | 118.48 | 227.31 | 150.36 | 10.83 | 7.56 |
| Mn | 310.81 | 176.67 | 87.44 | 73.86 | 5.32 | 7.70 |
| Cr | 110.70 | 241.50 | -36.50 | -55.77 | -4.02 | 16.95 |
| Ti | 137.10 | 98.10 | 55.80 | 31.60 | 2.28 | 3.13 |
| Mo | 120.90 | 305.10 | 13.70 | -56.83 | -4.09 | 15.14 |
| Si | 64.40 | 87.20 | 4.70 | -6.03 | -0.43 | 7.22 |
| Nb | -24.30 | 264.60 | -67.30 | -118.73 | -8.55 | 5.62 |

*Table 3. Ideal shear strength $\tau_{is}$ of fcc Ni and the corresponding engineering shear strain ε determined through DFT calculations with three different supercell sizes and nanoindentation measurement.*

| Slip system | Supercell size | ε | ${}^{0}\tau_{Ni,is}^{fcc}$ (GPa) | Method | Reference |
|---|---|---|---|---|---|
| $\{111\}\langle 11\bar{2}\rangle$ | 6-atom | 0.12 | 5.0 | Pure alias shear | [30] |
| $\{111\}\langle 11\bar{2}\rangle$ | 12-atom | 0.12 | 5.13 | Pure alias shear | [26] |
| $\{111\}\langle 11\bar{2}\rangle$ | 36-atom | 0.12 | 5.09 | Pure alias shear | This work |
| N/A | N/A | N/A | ~8 | Nanoindentation | [2] |

*Table 4. Ideal shear strength ($\tau_{is}$) summary of $Ni_{34}XZ$ determined by the first-principles calculations. The $\tau_{is}$ ranges from 4.16 to 5.24 GPa with a trend from highest to lowest when alloying elements into dilute Ni alloys, i.e., Co > Fe> Mn > Ni (i.e., pure Ni), > Cr > Al > Ti > Mo > Si > Nb.*

| Co | Fe | Mn | Ni | Cr | Al | Ti | Mo | Si | Nb | |
|---|---|---|---|---|---|---|---|---|---|---|
| 5.24 | 5.18 | 5.20 | 5.20 | 5.10 | 4.98 | 4.87 | 4.83 | 4.83 | 4.62 | Co |
| | 5.17 | 5.17 | 5.14 | 5.05 | 4.94 | 4.82 | 4.79 | 4.73 | 4.57 | Fe |
| | | 5.17 | 5.10 | 5.02 | 4.92 | 4.84 | 4.76 | 4.71 | 4.57 | Mn |
| | | | 5.09 | 5.03 | 4.85 | 4.77 | 4.74 | 4.68 | 4.51 | Ni |
| | | | | 4.88 | 4.88 | 4.70 | 4.64 | 4.73 | 4.51 | Cr |
| | | | | | 4.73 | 4.64 | 4.60 | 4.53 | 4.42 | Al |
| | | | | | | 4.57 | 4.52 | 4.41 | 4.35 | Ti |
| | | | | | | | 4.45 | 4.41 | 4.31 | Mo |
| | | | | | | | | 4.32 | 4.18 | Si |
| | | | | | | | | | 4.16 | Nb |

*Table 5. Binary interaction parameters ${}^{0}\phi_{Ni,j}^{fcc}$ (GPa)using the results of $Ni_{35}X$, $Ni_{34}X_2$ and $Ni_{11}X$.*

| Alloying element X | Co | Fe | Mn | Ni | Cr | Al | Ti | Mo | Si | Nb |
|---|---|---|---|---|---|---|---|---|---|---|
| ${}^{0}\phi_{Ni,j}^{fcc}$ (GPa) | 2.77 | -4.67 | 0.48 | 0 | 6.87 | -4.34 | -7.87 | 0.37 | -7.21 | -2.02 |

*Table 6. Ternary interaction parameters* ${}^{0}\phi_{Ni,j,k}^{fcc}(GPa)$.

| Co | Fe | Mn | Ni | Cr | Al | Ti | Mo | Si | Nb | |
|---|---|---|---|---|---|---|---|---|---|---|
| 0.00 | -78.56 | -26.55 | -0.58 | -38.04 | -53.68 | -62.79 | -171.61 | -38.17 | -203.80 | Co |
| | 0.00 | 33.27 | 18.08 | -5.66 | -7.58 | -30.41 | -125.51 | -74.41 | -171.42 | Fe |
| | | 0.00 | -12.25 | -22.27 | -10.46 | 21.60 | -142.11 | -77.29 | -146.86 | Mn |
| | | | 0.00 | 17.43 | -80.54 | -48.49 | -143.59 | -92.48 | -203.22 | Ni |
| | | | | 0.00 | 60.39 | -44.78 | -181.05 | 75.89 | -103.46 | Cr |
| | | | | | 0.00 | 21.92 | -86.90 | -49.52 | -77.93 | Al |
| | | | | | | 0.00 | -54.85 | -72.36 | -32.15 | Ti |
| | | | | | | | 0.00 | -126.29 | -140.97 | Mo |
| | | | | | | | | 0.00 | -185.93 | Si |
| | | | | | | | | | 0.00 | Nb |

## 11 References

[1] S. Ogata, J. Li, S. Yip, Ideal Pure Shear Strength of Aluminum and Copper, Science (1979) 298 (2002) 807–811. https://doi.org/10.1126/science.1076652.

[2] D. Lorenz, A. Zeckzer, U. Hilpert, P. Grau, H. Johansen, H.S. Leipner, Pop-in effect as homogeneous nucleation of dislocations during nanoindentation, Phys Rev B 67 (2003) 172101. https://doi.org/10.1103/PhysRevB.67.172101.

[3] D. Roundy, C.R. Krenn, M.L. Cohen, J.W. Morris, The ideal strength of tungsten, Philosophical Magazine A 81 (2001) 1725–1747. https://doi.org/10.1080/01418610108216634.

[4] C.R. Krenn, D. Roundy, M.L. Cohen, D.C. Chrzan, J.W. Morris, Connecting atomistic and experimental estimates of ideal strength, Phys Rev B 65 (2002) 134111. https://doi.org/10.1103/PhysRevB.65.134111.

[5] J.D. Shimanek, S. Qin, S.L. Shang, Z.K. Liu, A.M. Beese, Predictive Crystal Plasticity Modeling of Single Crystal Nickel Based on First-Principles Calculations, JOM (2022). https://doi.org/10.1007/s11837-022-05175-6.

[6] B. Joós, M.S. Duesbery, The Peierls Stress of Dislocations: An Analytic Formula, Phys Rev Lett 78 (1997) 266–269. https://doi.org/10.1103/PhysRevLett.78.266.

[7] V. Shankar, K. Bhanu Sankara Rao, S.L. Mannan, Microstructure and mechanical properties of Inconel 625 superalloy, Journal of Nuclear Materials 288 (2001) 222–232. https://doi.org/10.1016/S0022-3115(00)00723-6.

[8] M. Rahman, W.K.H. Seah, T.T. Teo, The Machinability of Incone1 718 Materials Processing Technology, 1997.

[9] L. Kumar, R. Venkataramani, M. Sundararaman, P. Mukhopadhyay, S.P. Garg, Studies on the oxidation behavior of Inconel 625 between 873 and 1523 K, Oxidation of Metals 45 (1996) 221–244. https://doi.org/10.1007/BF01046827.

[10] H. Pei, Z. Wen, Y. Zhang, Z. Yue, Oxidation behavior and mechanism of a Ni-based single crystal superalloy with single α-Al 2 O 3 film at 1000 °C, Appl Surf Sci 411 (2017) 124–135. https://doi.org/10.1016/j.apsusc.2017.03.116.

[11] S. Ma, Q. Ding, X. Wei, Z. Zhang, H. Bei, The Effects of Alloying Elements Cr, Al, and Si on Oxidation Behaviors of Ni-Based Superalloys, Materials 15 (2022) 7352. https://doi.org/10.3390/ma15207352.

[12] E. Anzini, N. Glaenzer, P.M. Mignanelli, M.C. Hardy, H.J. Stone, S. Pedrazzini, The effect of manganese and silicon additions on the corrosion resistance of a polycrystalline nickel-based superalloy, Corros Sci 176 (2020) 109042. https://doi.org/10.1016/j.corsci.2020.109042.

[13] A. Devaux, L. Nazé, R. Molins, A. Pineau, A. Organista, J.Y. Guédou, J.F. Uginet, P. Héritier, Gamma double prime precipitation kinetic in Alloy 718, Materials Science and Engineering: A 486 (2008) 117–122. https://doi.org/10.1016/j.msea.2007.08.046.

[14] A.J. Ardell, The growth of gamma prime precipitates in aged Ni−Ti alloys, Metallurgical Transactions 1 (1970) 525–534. https://doi.org/10.1007/BF02811564.

[15] C.T. Sims, A History of Superalloy Metallurgy for Superalloy Metallurgists, in: Superalloys 1984 (Fifth International Symposium), TMS, 1984: pp. 399–419. https://doi.org/10.7449/1984/Superalloys_1984_399_419.

[16] E. Akca, A. Gürsel, A Review on Superalloys and IN718 Nickel-Based INCONEL Superalloy, Periodicals of Engineering and Natural Sciences (PEN) 3 (2015). https://doi.org/10.21533/pen.v3i1.43.

[17] M. Detrois, P.D. Jablonski, S. Antonov, S. Li, Y. Ren, S. Tin, J.A. Hawk, Design and thermomechanical properties of a γ′ precipitate-strengthened Ni-based superalloy with high entropy γ matrix, J Alloys Compd 792 (2019) 550–560. https://doi.org/10.1016/j.jallcom.2019.04.054.

[18] K. Wang, B. Chang, Y. Lei, H. Fu, Y. Lin, Effect of Cobalt on Microstructure and Wear Resistance of Ni-Based Alloy Coating Fabricated by Laser Cladding, Metals (Basel) 7 (2017) 551. https://doi.org/10.3390/met7120551.

[19] Z.K. Liu, First-principles calculations and CALPHAD modeling of thermodynamics, J Phase Equilibria Diffus 30 (2009) 517–534. https://doi.org/10.1007/s11669-009-9570-6.

[20] J. Andersson, J. Ågren, Models for numerical treatment of multicomponent diffusion in simple phases, J Appl Phys 72 (1992) 1350–1355. https://doi.org/10.1063/1.351745.

[21] B. Hallstedt, N. Dupin, M. Hillert, L. Höglund, H.L. Lukas, J.C. Schuster, N. Solak, Thermodynamic models for crystalline phases. Composition dependent models for volume, bulk modulus and thermal expansion, Calphad 31 (2007) 28–37. https://doi.org/10.1016/j.calphad.2006.02.008.

[22] Z.K. Liu, H. Zhang, S. Ganeshan, Y. Wang, S.N. Mathaudhu, Computational modeling of effects of alloying elements on elastic coefficients, Scr Mater 63 (2010) 686–691. https://doi.org/10.1016/j.scriptamat.2010.03.049.

[23] M. Asadikiya, V. Drozd, S. Yang, Y. Zhong, Enthalpies and elastic properties of Ni-Co binary system by ab initio calculations and an energy comparison with the CALPHAD approach, Mater Today Commun 23 (2020) 100905. https://doi.org/10.1016/j.mtcomm.2020.100905.

[24] S. Shang, Y. Wang, Y. Du, M.A. Tschopp, Z.K. Liu, Integrating computational modeling and first-principles calculations to predict stacking fault energy of dilute multicomponent Ni-base alloys, Comput Mater Sci 91 (2014) 50–55. https://doi.org/10.1016/j.commatsci.2014.04.040.

[25] U.R. Kattner, THE CALPHAD METHOD AND ITS ROLE IN MATERIAL AND PROCESS DEVELOPMENT, Tecnol Metal Mater Min 13 (2016) 3–15. https://doi.org/10.4322/2176-1523.1059.

[26] J.D. Shimanek, S.-L. Shang, A.M. Beese, Z.-K. Liu, Insight into ideal shear strength of Ni-based dilute alloys using first-principles calculations and correlational analysis, Comput Mater Sci 212 (2022) 111564. https://doi.org/10.1016/j.commatsci.2022.111564.

[27] X. Chong, S.-L. Shang, A.M. Krajewski, J.D. Shimanek, W. Du, Y. Wang, J. Feng, D. Shin, A.M. Beese, Z.-K. Liu, Correlation analysis of materials properties by machine learning: illustrated with stacking fault energy from first-principles calculations in dilute fcc-based alloys, Journal of Physics: Condensed Matter 33 (2021) 295702. https://doi.org/10.1088/1361-648X/ac0195.

[28] S.L. Shang, J. Shimanek, S. Qin, Y. Wang, A.M. Beese, Z.K. Liu, Unveiling dislocation characteristics in N i3Al from stacking fault energy and ideal strength: A first-principles study via pure alias shear deformation, Phys Rev B 101 (2020). https://doi.org/10.1103/PhysRevB.101.024102.

[29] M. Jahnátek, J. Hafner, M. Krajčí, Shear deformation, ideal strength, and stacking fault formation of fcc metals: A density-functional study of Al and Cu, Phys Rev B Condens Matter Mater Phys 79 (2009). https://doi.org/10.1103/PhysRevB.79.224103.
[30] S.L. Shang, W.Y. Wang, Y. Wang, Y. Du, J.X. Zhang, A.D. Patel, Z.K. Liu, Temperature-dependent ideal strength and stacking fault energy of fcc Ni: A first-principles study of shear deformation, Journal of Physics Condensed Matter 24 (2012). https://doi.org/10.1088/0953-8984/24/15/155402.
[31] S.L. Shang, W.Y. Wang, B.C. Zhou, Y. Wang, K.A. Darling, L.J. Kecskes, S.N. Mathaudhu, Z.K. Liu, Generalized stacking fault energy, ideal strength and twinnability of dilute Mg-based alloys: A first-principles study of shear deformation, Acta Mater 67 (2014) 168–180. https://doi.org/10.1016/j.actamat.2013.12.019.
[32] T. Bučko, J. Hafner, J.G. Ángyán, Geometry optimization of periodic systems using internal coordinates, Journal of Chemical Physics 122 (2005). https://doi.org/10.1063/1.1864932.
[33] G. Kresse, J. Furthmüller, Efficient iterative schemes for ab initio total-energy calculations using a plane-wave basis set, Phys Rev B 54 (1996) 11169–11186. https://doi.org/10.1103/PhysRevB.54.11169.
[34] G. Kresse, D. Joubert, From ultrasoft pseudopotentials to the projector augmented-wave method, Phys Rev B 59 (1999) 1758–1775. https://doi.org/10.1103/PhysRevB.59.1758.
[35] J.P. Perdew, J.A. Chevary, S.H. Vosko, K.A. Jackson, M.R. Pederson, D.J. Singh, C. Fiolhais, Atoms, molecules, solids, and surfaces: Applications of the generalized gradient approximation for exchange and correlation, Phys Rev B 46 (1992) 6671–6687. https://doi.org/10.1103/PhysRevB.46.6671.
[36] M. Methfessel, A.T. Paxton, High-precision sampling for Brillouin-zone integration in metals, Phys Rev B 40 (1989) 3616–3621. https://doi.org/10.1103/PhysRevB.40.3616.
[37] Sk. SKRIEs, H.A. Mooed, Qtf jonrnal of enperimental and theoretical physics established by E. L Nichols in 1893 Magnetic Moment Distribution of Nickel Metal*, 1966.
[38] M.F. Collins, J.B. Forsyth, The magnetic moment distribution in some transition metal alloys, Philosophical Magazine 8 (1963) 401–410. https://doi.org/10.1080/14786436308211141.
[39] T. Kanomata, H. Ido, T. Kaneko, Effect of Pressure on Curie Temperature of Calcogenide Spinels $CuCr_2X_4$ (X=S, Se and Te), J Physical Soc Japan 29 (1970) 332–335. https://doi.org/10.1143/JPSJ.29.332.
[40] X. Liu, C.Z. Wang, H.Q. Lin, K.M. Ho, Magnetic moment enhancement for Mn7 cluster on graphene, Journal of Physical Chemistry C 118 (2014) 19123–19128. https://doi.org/10.1021/jp504329c.
[41] A. van de Walle, Multicomponent multisublattice alloys, nonconfigurational entropy and other additions to the Alloy Theoretic Automated Toolkit, CALPHAD 33 (2009) 266–278. https://doi.org/10.1016/j.calphad.2008.12.005.
[42] S. Shang, Y. Wang, W.Y. Wang, H. Fang, Z.K. Liu, Low energy structures of lithium-ion battery materials Li(Mn xNixCo1-2x)O2 revealed by first-principles calculations, Appl Phys Lett 103 (2013). https://doi.org/10.1063/1.4817763.
[43] Y. Wang, S. Curtarolo, C. Jiang, R. Arroyave, T. Wang, G. Ceder, L.Q. Chen, Z.K. Liu, Ab initio lattice stability in comparison with CALPHAD lattice stability, CALPHAD 28 (2004) 79–90. https://doi.org/10.1016/j.calphad.2004.05.002.

[44] G.B. Olson, Z.K. Liu, Genomic materials design: CALculation of PHAse Dynamics, Calphad 82 (2023) 102590. https://doi.org/10.1016/j.calphad.2023.102590.
[45] S. Yang, Y. Wang, Z.-K. Liu, Y. Zhong, Ab initio simulations on the pure Cr lattice stability at 0K: Verification with the Fe-Cr and Ni-Cr binary systems, Calphad 75 (2021) 102359. https://doi.org/10.1016/j.calphad.2021.102359.
[46] Z.K. Liu, Thermodynamics and its prediction and CALPHAD modeling: Review, state of the art, and perspectives, CALPHAD 82 (2023) 102580. https://doi.org/10.1016/j.calphad.2023.102580.
[47] S.-L. Shang, M.C. Gao, Y. Wang, J. Li, A.M. Beese, Z.-K. Liu, Mechanical properties of pure elements from a comprehensive first-principles study to data-driven insights, Materials Science and Engineering: A (2024) 147446. https://doi.org/10.1016/j.msea.2024.147446.
[48] K.M. Knowles, P.R. Howie, The Directional Dependence of Elastic Stiffness and Compliance Shear Coefficients and Shear Moduli in Cubic Materials, J Elast 120 (2015) 87–108. https://doi.org/10.1007/s10659-014-9506-1.
[49] S.L. Shang, A. Saengdeejing, Z.G. Mei, D.E. Kim, H. Zhang, S. Ganeshan, Y. Wang, Z.K. Liu, First-principles calculations of pure elements: Equations of state and elastic stiffness constants, Comput Mater Sci 48 (2010) 813–826. https://doi.org/10.1016/j.commatsci.2010.03.041.
[50] S.-L. Shang, B.-C. Zhou, W.Y. Wang, A.J. Ross, X.L. Liu, Y.-J. Hu, H.-Z. Fang, Y. Wang, Z.-K. Liu, A comprehensive first-principles study of pure elements: Vacancy formation and migration energies and self-diffusion coefficients, Acta Mater 109 (2016) 128–141. https://doi.org/10.1016/j.actamat.2016.02.031.
[51] M. Lee, M. Kim, K. Min, Evaluation of principal features for predicting bulk and shear modulus of inorganic solids with machine learning, Mater Today Commun 33 (2022) 104208. https://doi.org/10.1016/j.mtcomm.2022.104208.
[52] A.E. Isabelle Guyon, An Introduction to Variable and Feature Selection, Procedia Comput Sci 94 (2016) 465–472.
[53] M. ROBNIK SIKONJA MarkoRobnik, friuni-ljsi IGOR KONONENKO IgorKononenko, Theoretical and Empirical Analysis of ReliefF and RReliefF, Mach Learn 53 (2003) 23–69.
[54] The Mathworks Inc., MATLAB - MathWorks, Www.Mathworks.Com/Products/Matlab (2020).
[55] X. Chong, S. Shang, A.M. Krajewski, J. Shimanek, W. Du, Y. Wang, F. Jing, D. Shin, A.M. Beese, Z.-K. Liu, Correlation analysis of materials properties by machine learning: Illustrated with stacking fault energy from first-principles calculations in dilute fcc-based alloys, Journal of Physics: Condensed Matter (2021). https://doi.org/10.1088/1361-648X/ac0195.
[56] S.L. Shang, W.Y. Wang, Y. Wang, Y. Du, J.X. Zhang, A.D. Patel, Z.K. Liu, Temperature-dependent ideal strength and stacking fault energy of fcc Ni: a first-principles study of shear deformation, Journal of Physics-Condensed Matter 24 (2012) 155402. https://doi.org/155402 10.1088/0953-8984/24/15/155402.
[57] R.T. Sanderson, Handbook of Molecular Constants of Inorganic Compounds, Academic Press, 1983. https://pubs.acs.org/sharingguidelines.
[58] S. Ogata, J. Li, Y. Shibutani, S. Yip, Ab initio study of ideal shear strength, in: IUTAM Symposium on Mesoscopic Dynamics of Fracture Process and Materials Strength:

Proceedings of the IUTAM Symposium Held in Osaka, Japan, 6--11 July 2003, 2004: pp. 401–410.
[59] O. Redlich, A.T. Kister, Algebraic Representation of Thermodynamic Properties and the Classification of Solutions, Ind Eng Chem 40 (1948) 345–348. https://doi.org/10.1021/ie50458a036.
[60] L. Kaufman, H. Bernstein, Computer calculation of phase diagrams With special reference to refractory metals, Academic Press Inc, United States, 1970. http://inis.iaea.org/search/search.aspx?orig_q=RN:02004171.
[61] Q. Chen, B. Sundman, Calculation of debye temperature for crystalline structures—a case study on Ti, Zr, and Hf, Acta Mater 49 (2001) 947–961. https://doi.org/10.1016/S1359-6454(01)00002-7.
[62] O.L. Anderson, A simplified method for calculating the debye temperature from elastic constants, Journal of Physics and Chemistry of Solids 24 (1963) 909–917. https://doi.org/10.1016/0022-3697(63)90067-2.